\documentstyle[floats,epsf,prl,aps,twocolumn]{revtex} 
\begin{document} 
\frenchspacing 
 
\wideabs{
\title{Commensurability effects in Andreev antidot billiards}
\author{J.\ Eroms\cite{adresse},  M.\ Tolkiehn, D.\ Weiss, U. R\"ossler} 
\address{Universit\"at Regensburg, D-93040 Regensburg, Germany,} 
\author{J.\ De Boeck, G.\ Borghs} 
\address{IMEC, Kapeldreef 75, B-3001 Leuven, Belgium} 
\date{\today} 
\maketitle 
\begin{abstract} 
An Andreev billiard was realized in an array of niobium filled antidots in a high-mobility InAs/AlGaSb heterostructure. Below the critical temperature $T_C$ of the Nb dots we observe a strong reduction of the resistance around $B=0$ and a suppression of the commensurability peaks, which are usually found in antidot lattices. Both effects can be explained in a classical Kubo approach by considering the trajectories of charge carriers in the semiconductor, when Andreev reflection at the semiconductor-superconductor interface is included. For perfect Andreev reflection, we expect a complete suppression of the commensurability features, even though motion at finite $B$ is chaotic.
\end{abstract} 
} 
 
Antidot lattices, consisting of a periodic array of minuscule voids in a two-dimensional electron gas (2DEG) constitute a model system to study the classical dynamics of electrons in a periodic potential, if the periods are larger than the Fermi wavelength but much smaller than the electron mean free path\cite{Antidots,andospringer}. The electrons move with the Fermi velocity between the repulsive potential pillars which get formed when, e.g., a periodic array of holes is etched into the top layers of a semiconductor heterostructure by means of conventional nanofabrication. Both at zero and finite magnetic field the electrons get specularly reflected at the potential posts and the resulting motion is chaotic as is illustrated in Fig.~1a. The antidot lattice can be considered as a periodically repeated Sinai billiard. The dynamics of the electrons is reflected by peaks in the resistance, measured as a function of the applied magnetic field $B$ and shown in Fig.~1c. Peaks emerge whenever the classical cyclotron diameter $2R_C$ of the electrons at the Fermi energy fits around a certain number of antidots, while the zero field resistance of an antidot array is markedly higher than the one of an unpatterned 2DEG. The resistance drops at higher magnetic fields once the cyclotron orbit of the electrons becomes smaller than the distance between adjacent antidots. The usual way to model magnetoresistance traces like the one in Fig.~1c is to calculate numerically 
the electron trajectories in a model antidot potential and to evaluate the velocity correlation function within a classical Kubo approach\cite{Ragnar}. The structure in the magnetoresistance in this picture reflects correlations in the chaotic sea.

Here we explore a system where the specular reflection at the potential posts is, at least partly, replaced by Andreev reflection\cite{Andreev} with an incident electron retro-reflected as a hole. This has been achieved by placing an array of superconducting niobium disks on an InAs quantum well containing a 2DEG. While in a perfectly Andreev reflecting antidot array the trajectories of electrons and retro-reflected holes behave regularly at $B=0$, their dynamics becomes again chaotic at finite $B$\cite{chaos} as is illustrated in Fig.~1d. The system can be considered as an experimental version of an Andreev billiard whose properties are currently of considerable theoretical interest\cite{chaos,beenakker,richter}. An early experimental attempt of realizing a similar system was reported by Rahman et. al.\cite{rahman}. Here we address the question how retroreflection, instead of specular reflection, modifies the magnetoresistance features of an antidot array.

%
\begin{figure}[htb]\vspace{-0.8em} \begin{minipage}{8.6cm} \epsfxsize=8.6cm \epsfbox{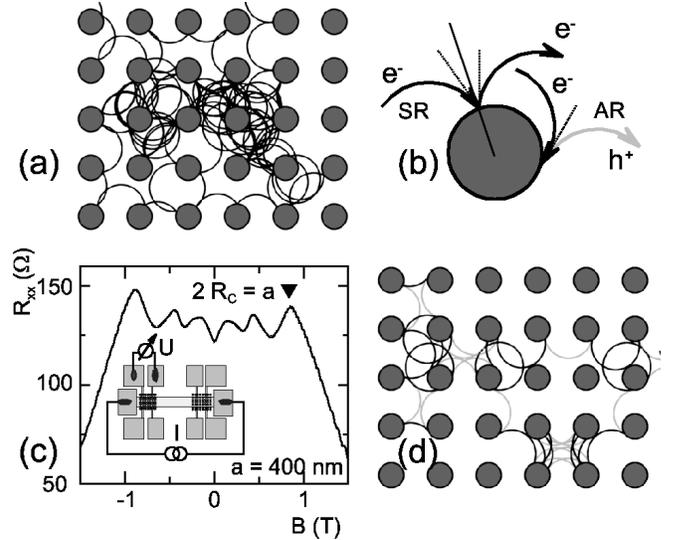} \end{minipage} \vspace{0.2cm} \caption{Chaotic trajectories calculated at $2R_C=a$ in specularly reflecting (a) and Andreev reflecting (d) antidot lattices.  (c): Magnetoresistance of an antidot lattice taken at $T=1.5$\ K. The inset shows the Hall bar geometry and measurement setup. (b): Trajectories of specularly (SR) and Andreev reflected (AR) particles  Black (Gray) trajectory sections: electrons (holes)} \label{motivation}\vspace{-1em}\end{figure} 

\begin{figure}[!htb]\vspace{-0.8em}\begin{minipage}{8.6cm} \epsfxsize=8.6cm \epsfbox{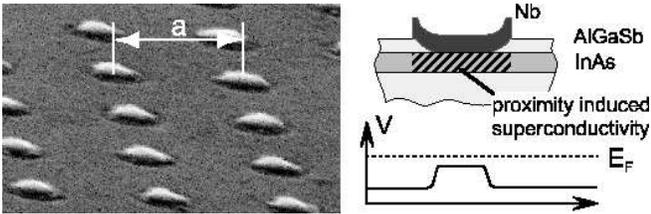} \end{minipage} \vspace{0.2cm} \caption{Left: The SEM-picture taken at an angle of $60^{\rm o}$ shows Nb filled antidots with a lattice period $a$ of 400 nm. Right: A cross-sectional view of the sample design. The InAs underneath the Nb dot is superconducting due to the proximity effect. The electrostatic antidot potential is sketched below.} \label{sample}\vspace{-1em}\end{figure}

We prepared samples with square lattices of niobium filled antidots in an InAs-based 2DEG structure (see Fig.\ \ref{sample}). 
Due to the lack of a Schottky-barrier, InAs is the material of choice for highly transparent semiconductor-superconductor contacts \cite{InAs,nguyen,morpurgo}. The lattice periods ranged between $a=300$\ nm and $a=1500$\ nm, and the ratio of antidot diameter and lattice period $d/a$ was varied between $0.20$ and $0.75$, with a number of different diameters for each lattice period.
The observations described in this article were most clearly visible at small $d/a$-values.

%
%

The samples were fabricated from a high-mobility InAs/Al$_{0.30}$Ga$_{0.70}$Sb heterostructure. The InAs-layer containing the 2DEG was 15\ nm thick and the top barrier layer of Al$_{0.30}$Ga$_{0.70}$Sb had a thickness of 10\ nm \cite{mbe}.   
Hall bars were prepared employing photolithography and wet chemical etching and the 2DEG was contacted with non-alloyed Cr/Au pads. At this stage, the electron density in the 2DEG was $n_s=1.25\times 10^{12}$\ cm$^{-2}$ with an electron mobility of $\mu=200\,000$\ cm$^2$/Vs at $T=1.5$\ K, resulting in a mean free path of about $3.8\ \mu$m.  
The antidot lattices were defined by electron beam lithography and a selective wet chemical etch removing only the top AlGaSb layer at the antidot sites. Subsequently, the niobium disks were deposited with DC magnetron sputtering and lift off. 
Both patterning steps were carried out using the same PMMA mask. In this way a complicated realignment was avoided and, more importantly, the 
 exposure to air after etching was minimized. To compensate for the inevitable underetching of the PMMA mask by the wet etching step, a modified developing scheme was used \cite{ms2000}.
This procedure proved to be crucial for a highly transparent contact between the 2DEG and the superconductor.
The samples were etched and developed immediately before loading them into the load-lock chamber of the sputtering system. Despite the short exposure to air it was necessary to remove residual contamination on the InAs surface with a low power Argon sputter etch \cite{InAs}. By this technique, high quality samples could be routinely fabricated, which consistently displayed similar experimental results. Unpatterned niobium films prepared in the same system had a critical temperature $T_C$ of about 8 K and a critical field $B_C$ of about 2.4\ T at 1.5\ K. Niobium dots with a minimum diameter of 170\ nm were still superconducting with a $T_C$ of about 5\ K.

The measurements were performed in a helium cryostat with a variable temperature insert and a superconducting magnet system. Standard lock-in techniques were used to record the magnetoresistance traces. To minimize rf coupling to the samples, all leads were filtered with $\pi$-feedthrough filters on top of the cryostat.

\begin{figure}[htb] \begin{minipage}{8.6cm} \epsfxsize=8.6cm \epsfbox{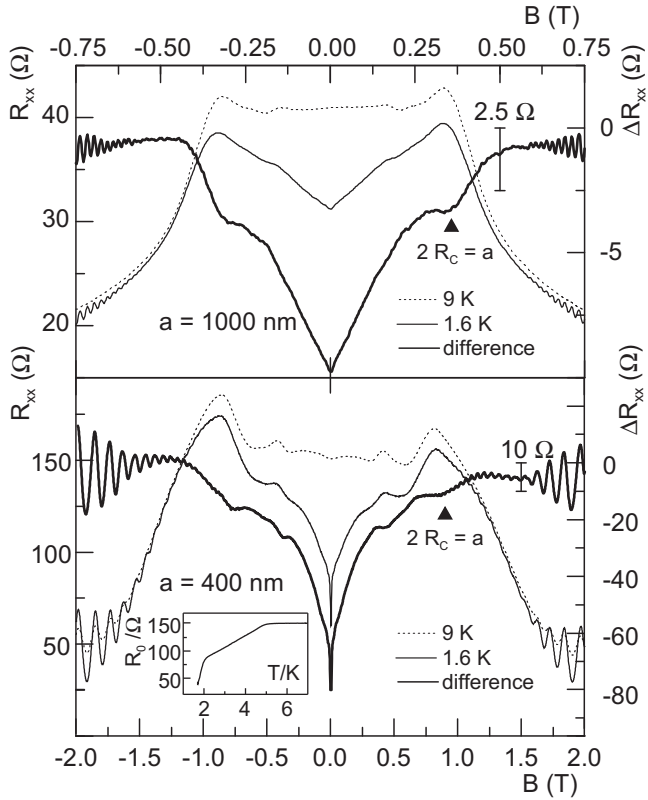} \end{minipage} \vspace{0.2cm} \caption{Magnetoresistance curves for samples with two different lattice periods $a$. In the upper graph, $a=1000$\ nm and the diameter $d=200$\ nm. The difference between the curves below $T_C$ and above $T_C$ shows a clear dip at $2R_C=a$. Lower graph: $a=400$\ nm and $d=170$\ nm. Two dips are visible for magnetic fields at which the antidot graph ($T=9$\ K) shows peaks. Inset: The resistance at $B=0$ starts to drop at $T=5$\ K and is reduced more sharply below $T=2$\ K, when the Nb disks begin to get Josephson coupled.} \label{experiment}\vspace{-1em}\end{figure}

The dashed graphs in Fig.~\ref{experiment}a and \ref{experiment}b show data taken at 9 K, above $T_C$ of the niobium dots. For both samples the fundamental antidot peaks at $2R_C=a$ are clearly visible. The sample with the smaller period of $a=400$\ nm displays even more structure at lower $B$, ascribed to higher order commensurate orbits. From the small amplitude of the commensurability peaks we conclude that the 2DEG underneath the Nb disks is not completely devoid of electrons and that hence the repulsive potential maxima do not perforate the Fermi energy, as sketched in Fig.\ \ref{sample}, lower right\cite{aetztiefe}. Thus a fraction of the electrons in phase space is able to explore the 2DEG underneath the Nb disks. The traces change substantially if cooled below $T_C$. The thin solid lines in Fig.~\ref{experiment}  were taken at 1.5\ K and show a magnetic field dependent reduction of the resistance which is especially pronounced at low $B$. The `V-shaped' magnetoresistance around $B=0$ is characteristic for good Nb/2DEG contacts and is not observed in devices with lower quality contacts. We ascribe this behaviour to Andreev reflection. More precisely, we assume that Andreev reflection is taking place between the normal 2DEG and a region of proximity induced superconductivity in the InAs underneath the Nb disks \cite{merkt,kuemmel}.
In the $B$-field region where the commensurability peaks are observed, reflection is mostly specular, and consequently, the antidot peaks are still present in the 1.5\ K data.

In the millitesla range and at temperatures below 2\ K (see inset in Fig.~\ref{experiment}), the resistance drops again sharply, the niobium dots are Josephson coupled and we observe oscillations (not shown) with dips at integer and fractional multiples of one flux quantum $\Phi_0=h/2e$ per unit cell \cite{jja}. 
At $T=0.5$\ K, the whole array becomes superconducting up to fields of several millitesla. In this article, however, only the classical features visible at higher temperatures and magnetic fields are dealt with. 
%
%

To extract the influence of Andreev reflection on the commensurability features we subtracted the above-$T_C$-trace from the below-$T_C$-trace. 
The result is shown as $\Delta R_{xx}$ in Fig.\ \ref{experiment} a, b.
Surprisingly, $\Delta R_{xx}$  exhibits dips at commensurate magnetic fields where the original traces display maxima. In contrast, in conventional density modulated electron systems the amplitudes of the commensurability features grow with decreasing temperature due to the reduced broadening of the Fermi distribution. This would result in peaks of $\Delta R_{xx}$ instead of dips. 
To confirm that the observed dips are indeed due to commensurability effects, we prepared a reference sample with Nb disks distributed randomly on the InAs layer. As expected, this sample shows no commensurability features but a monotonously decreasing $\Delta R_{xx}$ only. The dips observed in $\Delta R_{xx}$ hence indicate a suppression of commensurability peaks. We will show below that such a suppression is expected for electrons being Andreev reflected from the antidots. In case of pure Andreev scattering (no specularly reflected electrons), surprisingly, all commensurability features are predicted to disappear.

To show this we carried out magnetotransport calculations employing the classical Kubo formula, which previously gave excellent agreement with the experimental data for the purely electrostatic lateral superlattice\cite{Ragnar}. For the present situation we modify this formula for the components $\sigma_{ij}$ of the velocity tensor:

\begin{equation} 
\sigma_{ij}=\frac{m^*}{\pi \hbar^2} 
\int_0^\infty \!\!\!\!{\rm d}t\, e^{-t/\tau}\langle q(t)v_i(t)\,q(0)v_j(0)\rangle 
\label{Kubo}
\end{equation} 
by using a time-dependent charge $q(t)$, which is $+e$ ($-e$) on the hole (electron) part of the trajectory.
Impurity scattering is included by the damping factor $e^{-t/\tau}$.
Thus, in our description Andreev reflection---which is a quantum effect that can be described using the Bogliubov-de Gennes equations---is only taken into account by the resulting classical electron and hole trajectories in the normal region of the 2DEG. We do not account for transport through the metallic or superconducting dots. Since the mean free path in the samples is much longer than the lattice period, the superconducting regions do not provide a transport channel with a much lower resistance than the ballistic 2DEG. As long as the Nb dots are not Josephson coupled, the current is carried by normal state electrons and holes in the 2DEG.

%
%
\begin{figure}[htb] \begin{minipage}{8.6cm} \epsfxsize=8.6cm \epsfbox{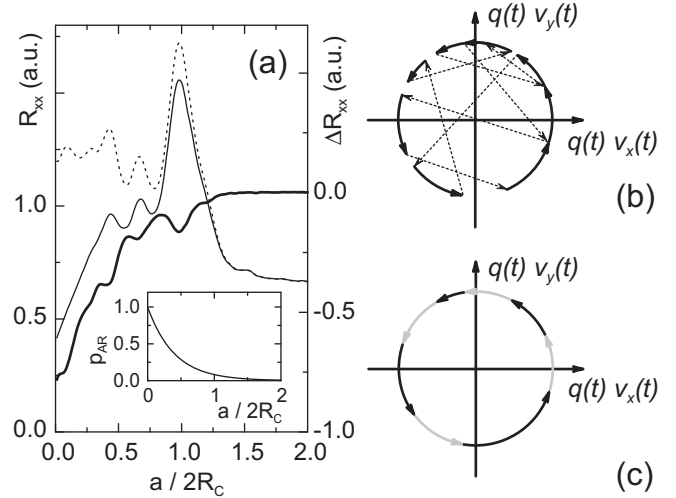}\end{minipage} \vspace{0.2cm} \caption{(a): Calculated magnetoresistance curves for $p_{AR}=0$ (dashed line) and a magnetic field dependent $p_{AR}$ (thin solid line). The difference of both graphs (thick solid line) has dips at the commensurability positions. Inset: Exponential dependence of $p_{AR}$ on $B$ used for the calculations. (b): Sketch of the velocity components of a trajectory in an antidot lattice with specular reflection. (c): The same, but with Andreev reflecting boundaries. Black: electrons, Gray: holes.}\label{theory}\vspace{-1em} \end{figure}

For perfect Andreev reflection, the Kubo formula yields a surprisingly simple result: Since all the velocity components are reversed upon Andreev reflection, 
the velocity correlation function $\langle q(t)v_i(t)\,q(0)v_j(0)\rangle$ of an unpatterned 2DEG is recovered. 
This is illustrated in Fig.~\ref{theory} where the velocity components $v_x$ and $v_y$ are compared for the case of pure specular (\ref{theory}b) and pure Andreev reflection (\ref{theory}c). In between reflection events the values of the velocity components move along a circle with the Fermi velocity $v_F$. In the case of perfect Andreev reflection both velocity components are reversed
but $q(t) v_x$ and $q(t) v_y$ continue to move along the circle since the sign of the charge $q(t)$ is also reversed (conversion of holes into electrons and vice versa). Hence the correlation function of a perfect Andreev antidot billiard is {\em identical\/} to the correlation function of the free electron motion. In sharp contrast, reflection on a conventional antidot boundary gives different velocity correlations depending on the exact initial conditions of the trajectory (Fig.~\ref{theory}b). 

In our samples scattering will be due to a mixture of both processes.
We therefore take into account an electrostatic scattering mechanism giving rise to specular reflection, i.e. we model the interface by a hard-wall potential. If the trajectory hits an antidot, the charge carrier is either Andreev reflected or specularly reflected with a certain probability.
The Andreev reflection probability $p_{AR}$ was not assumed to depend on the angle of incidence on the boundary since this should have little effect on the transport coefficients: In the extreme case of grazing incidence, where Andreev reflection is most strongly suppressed \cite{winkel}, the contribution to the conductivity is the same for Andreev reflection and normal reflection.
A set of magnetotransport curves was obtained by varying $p_{AR}$ from zero to one. For $p_{AR}=0$, the behaviour of a conventional antidot lattice is reproduced, whereas $p_{AR}=1$ gives a much lower resistance independent of $B$ just as for an unpatterned 2DEG.
Andreev reflection therefore provides a mechanism to explain both the reduced zero-field resistivity and the suppression of commensurability features on a common footing.

To explain the magnetic field dependence of the resistance below $T_C$, e.g. the V-shaped magnetoresistance, we have to assume a decrease of the Andreev reflection probability $p_{AR}$ with increasing magnetic field. While {\em a priori\/} not justified, this might be a consequence of a finite pair-amplitude
in the superconducting regions of the InAs \cite{magpar,vanson,kuemmel},
which decreases with increasing $B$. Within this picture,
$p_{AR}$ is affected by the pair-amplitude inside the 2DEG, which depends on temperature and magnetic field in a complicated way. No detailed explanation is presently available for this dependence. 

To compare with experiment we introduced phenomenologically an Andreev reflection probability $p_{AR}$ decreasing exponentially with $B$ and plotted in the inset of Fig.\ \ref{theory}a. While $p_{AR}$ is at maximum at $B=0$, reflection off the antidots is mostly specular at higher magnetic fields, where $p_{AR}$ is small. This explains why the commensurability peaks are not suppressed completely, but only partly.
The result of a corresponding model calculation is shown as a thin solid line in Fig.\ \ref{theory}a. The difference to the specularly reflecting antidot trace ($p_{AR}=0$, above $T_C$-trace) is shown as a thick solid line and shows the same dips, characteristic for the suppression of commensurability peaks, as observed in experiment. The result supports our assumption that the suppression of commensurability features is a consequence of Andreev reflection and the resulting peculiar electron-hole trajectories between antidots.

We emphasize that these findings are independent of the exact dependence of $p_{AR}$ on $B$. As long as there is a small degree of Andreev reflection present, the commensurability peaks are damped. Note that the calculated curves show more structure than the experimental ones. This is due to the fact that we used a hard wall potential at the antidot edges and also did not allow for fluctuations in the antidot diameter $d$. Nevertheless, our simple model seems to cover the essential physics.

In conclusion we have fabricated antidot lattices with Andreev reflecting boundaries in an InAs-based heterostructure. We observe antidot peaks above $T_C$ of the niobium dots and a suppression of commensurability maxima below $T_C$, which we can ascribe to the retroreflection of the charge carriers at the InAs-Nb-interfaces. This was confirmed by transport calculations based on the Kubo formula taking into account only the motion of electrons and holes in the 2DEG.

The authors would like to thank G. Bayreuther for the possibility to use the UHV sputtering system in his group for the niobium deposition and J. Keller, C. Strunk and R. K\"ummel for valuable discussions. Financial support by the Deusche Forschungsgemeinschaft (Graduiertenkolleg GRK 638) is gratefully acknowledged.

\newpage


\begin{thebibliography}{99}
\bibitem{aetztiefe} In control experiments antidot arrays with holes etched through the InAs quantum well showed much stronger commensurability features than those where the etching was only through the the AlGaSb top layer.
\end{thebibliography}
\end{document}